\documentclass[english,aps,prl,twocolumn,nofootinbib,floatfix,superscriptaddress,10pt]{revtex4-2}
\usepackage{amsmath,amssymb,braket,bbold,mathtools,graphicx,enumitem}
\usepackage[export]{adjustbox}
\usepackage[dvipsnames]{xcolor}
\usepackage[unicode=true,pdfusetitle, bookmarks=true,bookmarksnumbered=false,bookmarksopen=false, breaklinks=false,pdfborder={0 0 0},pdfborderstyle={},backref=false,colorlinks=true]
 {hyperref}
\hypersetup{colorlinks,allcolors=blue}
\usepackage[capitalize]{cleveref}

\usepackage[caption=false]{subfig}

\newcommand{\phantomsubfloat}[1]{\subfloat[][\hspace{0pt}]{#1}}
\newcommand{\phantomsubfloats}[1]{#1\vspace{-\baselineskip}}

\newcounter{subpanel}

\begin{document}

\title{Multibath Influence Matrices: Universal Scaling from Real-Time Dynamics}
 
\author{Matan Lotem}\affiliation{Department of Physics, Princeton University, Princeton, NJ 08544, USA}
\author{Michael Sonner}\affiliation{Department of Physics, University of California, Berkeley CA 94720, USA}
\author{Valentin Link}\affiliation{Institut für Physik und Astronomie, Technische Universität Berlin, D-10623, Berlin, Germany}

\author{Alessio Lerose}
\affiliation{Rudolf Peierls Centre for Theoretical Physics, University of Oxford, Oxford OX1 3NP, United Kingdom
}
\affiliation{Institute for Theoretical Physics, KU Leuven, Celestijnenlaan 200D, 3001 Leuven, Belgium}

\author{Dmitry A. Abanin}\affiliation{Department of Physics, Princeton University, Princeton, NJ 08544, USA}
\affiliation{\'{E}cole Polytechnique F\'{e}d\'{e}rale de Lausanne (EPFL), 1015 Lausanne, Switzerland}

\begin{abstract}
Diverging timescales are the hallmark of critical dynamics and the bottleneck
to their classical and quantum simulation. To tame this, we compress a
spacetime tensor network: temporally into semigroup influence matrices and
spatially via matrix-product states. Benchmarking on the two-impurity
Anderson model with its four fermionic species, we compute spectral
functions to map the evolution from a Kondo resonance, across a non-Fermi-liquid quantum critical point, and into a gapped singlet phase. Resolving transient through asymptotic dynamics in sudden quenches and Kibble-Zurek ramps, we obtain the
universal two-channel-Kondo exponent. 
Together, these results establish multibath influence matrices as a practical tool for real-time dynamics in strongly correlated multi-orbital systems.
\end{abstract}
 
\maketitle

\begin{figure}
    \vspace{-2em}
    \phantomsubfloats{
        \phantomsubfloat{\label{fig:model}}
        \phantomsubfloat{\label{fig:unitary_to_im}}
        \phantomsubfloat{\label{fig:mps_evo}}
    }
    \includegraphics[width=1\linewidth]{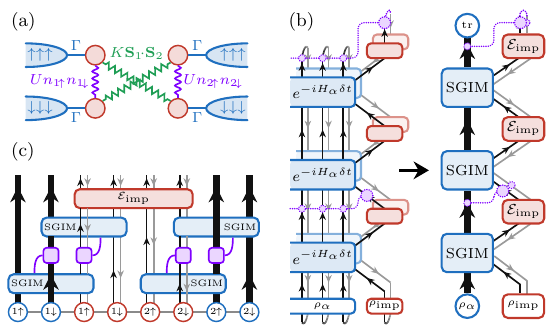}
    \caption{(a) The two-impurity Anderson model with four fermionic baths, interimpurity exchange, and intraimpurity Coulomb interactions. (b) Single-bath unitary dynamics (left) map to an influence matrix and impurity channel (right) with operator insertions and trailing Jordan-Wigner strings for a propagator calculation (purple). (c) MPS and Trotter decomposition of the full dynamical map with single-site parity MPOs (purple) attached to the SGIMs (blue) and a four-site impurity channel (red).
    }\vspace{-2em}
\end{figure}

Real-time dynamics directly probes exotic critical points of quantum impurity
models~\cite{nozieres1980Kondo,andrei1984Solution,tsvelick1985Thermodynamics,affleck1993Exact,grobis2007Kondo}
via critical slowing down~\cite{hertz1976Quantum,hohenberg1977Theory}, which
manifests in divergent quench relaxation
times~\cite{ratiani2010Nonequilibrium,mitra2018Quantum} and in the generation of
excitations when ramping a parameter across the transition, following the
Kibble-Zurek (KZ)
mechanism~\cite{kibble1976Topology,zurek1985Cosmological,ma2025Quantum}. These
dynamical signatures are natural probes of continuous phase transitions in cold-atom
experiments~\cite{braun2015Emergence,keesling2019Quantum,horowicz2021Critical,wu2024Indication,zhang2025Probing},
where impurity models have recently been
realized~\cite{bauer2013Realizing,wenz2013Few,riegger2018Localized,hewitt2024Controlling,zhang2020Controlling}.
In this Letter, we focus on the dynamics of the two-impurity Anderson model (2IAM) depicted
in \cref{fig:model}, where two interacting spinful quantum dots (impurities) are
independently Kondo-screened~\cite{kondo1964Resistance,hewson1997Kondo} by their
own noninteracting fermionic baths. Competition between this screening and a
direct interimpurity exchange gives rise to a two-channel Kondo (2CK)
non-Fermi-liquid critical point with fractional
excitations~\cite{jones1987Study, jones1988LowTemperature, jones1989Critical,
affleck1995Conformalfieldtheory, zarand2006Quantum,
mross2008Twoimpurity,sela2011Exact, mitchell2012TwoChannel}. 
Such critical points have already been observed in mesoscopic devices~\cite{potok2007Observation,roch2009Observation,keller2015Universal,iftikhar2015Twochannel,iftikhar2018Tunable,pouse2023Quantum,piquard2023Observing}, motivating dynamical protocols to probe nonequilibrium thermodynamics~\cite{saira2012Test,hofmann2016Equilibrium,hofmann2017Heat,barker2022Experimental,han2025Measuring}.
Besides probing
criticality, multibath impurity dynamics lies at the heart of
real-time dynamical mean-field theory
(DMFT)~\cite{georges1996Dynamical,aoki2014Nonequilibrium,kotliar2006Electronic}.

Simulating these dynamics numerically remains a formidable
challenge, highlighted by recent efforts to mitigate the sign problem in
quantum Monte
Carlo~\cite{eidelstein2020Multiorbital,erpenbeck2024Steadystate}
and the scaling of bond dimensions with the number of environments in tensor networks~\cite{bauernfeind2017Fork,cao2021Tree,fux2023Tensor,sun2025Scalable,wolf2015ImaginaryTime,stadler2015Dynamical,stadler2016Interleaved,grundner2024Complex,kugler2020Strongly}.
Accessing criticality, in particular, requires long evolution times at low temperature.
This is where methods based on the
\textcite{feynman2000Theory} influence functional excel, compressing a bath's
entire back-action on the impurity into a temporal matrix-product state (MPS)
with a bond dimension set by the range of temporal
correlations~\cite{strathearn2018Efficient,pollock2018NonMarkovian,lerose2021Influence,thoenniss2023Nonequilibrium,thoenniss2023Efficient,ng2023Realtime,chen2024Grassmann,park2024Tensor,fux2023Tensor,link2024Open,sun2025Scalable,keeling2026Process}.
For time-translation-invariant environments, using an MPS representation with a single repeated tensor, the
semigroup influence matrix (SGIM), achieves remarkable accuracy and extends accessible time scales~\cite{sonner2025Semigroup}. Here, we construct a multibath SGIM framework by
combining three techniques illustrated in
\cref{fig:mps_evo}: (i) temporal compression reduces each bath to an independent
charge-conserving SGIM, (ii) automated superfermion stitching combines these
into a single dynamical map, enforcing global fermionic anticommutation, and
(iii) spatial MPS compression of the joint state exploits low interbath
entanglement. By using high-accuracy representations of the individual SGIM and exploiting charge conservation through symmetric tensors, we demonstrate the full capabilities of the framework.

We apply it to the 2IAM, varying the interimpurity
exchange $K$ across the transition. The spectral function shows a resonance peak in the Kondo phase, which flattens at the critical point, giving way to a gapped spectral density. Resolving transient through asymptotic dynamics of a sudden
quench reveals the relaxation time divergence at the transition
$\tau\sim|K{-}K_c|^{-2}$ with the 2CK exponent. Remarkably, collapsing $\tau$
over temperature and bond dimension recovers the zero-temperature relaxation
across the full parameter range. Ramping $K$ at velocity $v$, the dissipated
work crosses over from a nonuniversal regime, through the universal KZ scaling
$v^{2/3}$, to a temperature-limited adiabatic one. Reading off these universal
scalings from real-time dynamics establishes the robustness of multibath
SGIM.


\paragraph{Multibath setup:} Splitting a quantum system into large environments and local impurities, we make a single structural
assumption: the environments are mutually independent, each coupling to the
impurities but not to other environments.
For the simplest case of time-independent Hamiltonians,
\begin{equation}\label{eq:H}
H=H_\mathrm{imp}+\sum_\alpha H_\alpha,
\end{equation}
where $H_\mathrm{imp}$ is the local impurity Hamiltonian, the sum runs over environments labeled by $\alpha$, and each $H_\alpha$ is supported on the impurity and environment $\alpha$.
As depicted in \cref{fig:unitary_to_im} for one environment (the 2IAM has four), the evolution along the forward ($+$) and backward ($-$) branches of the Keldysh contour Trotter decomposes as\footnote{When
$[H_\alpha,H_\alpha']=0$, as in the 2IAM, the product over environments is exact, and a
second-order Trotter decomposition is implemented at negligible cost.}
\begin{equation}
  \!\!\rho(t{+}\delta t) = e^{-iH\delta t}\rho(t)e^{iH\delta t},\ \ 
    e^{iH \delta t} \!\approx\! e^{iH_\mathrm{imp}\delta t} \!\prod_\alpha \!e^{iH_\alpha \delta t}\!.\!
\end{equation}
Starting from factorized initial density matrices $\rho(0)=\rho_\mathrm{imp}\bigotimes_\alpha\rho_\alpha$, correlated states are generated via evolution. After $t/\delta t$ steps, the Keldysh branches are joined to trace out the environments, yielding $\rho_\mathrm{imp}(t)$ or, with operator insertions and a final trace (purple), impurity correlators.
We emphasize that the framework introduced in this Letter is not restricted to Hamiltonian dynamics, and directly applies to interacting Floquet
baths~\cite{lerose2021Influence,giudice2022Temporal} as well as
dissipative~\cite{sonner2025Semigroup} and time-dependent
\cite{fux2021Efficient,kahlert2024Simulating,mickiewicz2026Exact} impurity
terms. Crucially, we allow strong impurity-environment coupling and do not
assume the environments act as Markovian baths. For the 2IAM, the environments are
Gaussian
\begin{equation}\label{eq:H_Gaussian} H_\alpha = \sum_k
 (v_{\alpha k}^{\phantom{\dagger}}\, d_{\alpha\phantom{k}\!\!\!}^\dagger c_{\alpha k}^{\phantom{\dagger}}+\mathrm{h.c.}) + \sum_k
 \varepsilon_k\, c_{\alpha k}^\dagger c_{\alpha k}^{\phantom{\dagger}}, \end{equation} with impurity
 $d_\alpha$ and environment $c_{\alpha k}$ fermionic operators, in an initial
 thermal state $\rho_\alpha\propto e^{-\beta \sum_k \varepsilon_k^{\phantom{\dagger}}\, c_{\alpha
 k}^\dagger c_{\alpha k}^{\phantom{\dagger}}}$, and bath memory encoded by the hybridization functions
 $\Gamma_\alpha(\omega)=\sum_k\pi|v_{\alpha k}|^2\delta(\omega-\varepsilon_k)$.


\paragraph{Temporal compression:}  Computing time-evolution of the impurity-environments system in \cref{fig:unitary_to_im} via an MPS representation of the global state is limited by the growth of spatial
 entanglement~\cite{calabrese2005Evolution, schuch2008Entropy}.
 Contracting instead all the environment degrees of freedom first (blue tensors) yields a `temporal state' known as
 an influence matrix~\cite{lerose2021Influence} or process
 tensor~\cite{pollock2018NonMarkovian}.
 Often, the temporal correlations encoded by this state are limited after the forward and backward degrees of freedom are folded together~\cite{banuls2009Matrix,hastings2015Connecting,lerose2023Overcoming}
 and it can be efficiently represented as a
 temporal MPS \cite{lerose2021Influence,sonner2021Influence,giudice2022Temporal,vilkoviskiy2024Bound,thoenniss2025Efficient}.
 For fermionic Gaussian baths as in \cref{eq:H_Gaussian}, different strategies for
 constructing such MPS have been proposed~\cite{thoenniss2023Efficient,
 ng2023Realtime, chen2024Grassmann, sonner2025Semigroup}.

We follow Ref.~\cite{sonner2025Semigroup}, which exploits time-translation
invariance of the environments and infinite MPS methods to obtain a
representation with a single repeated tensor, or semigroup influence matrix
(SGIM). It allows the formal replacement of the impurity-bath evolution with an
effective dissipative map that acts on a compressed-environment space, defined via the bond-space of the temporal MPS
\begin{equation}
    \includegraphics[valign=m]{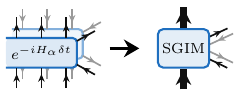}.
\end{equation}
The MPS structure naturally fuses both branches of the Keldysh contour into the bond, while the impurity carries them as separate forward and backward indices. Methods such as HEOM (hierarchical equations of motion)~\cite{tanimura2020Numerically} and pseudomode approaches~\cite{xu2026Colloquium,dorda2014Auxiliary,park2024QuasiLindblad} can be framed analogously~\cite{link2024Open,ortega-taberner2024Unifying}, with each environment replaced by a finite auxiliary space, albeit with more structure. Thus, our multibath stitching and spatial compression below are equally applicable to these approaches, as well as to influence matrices in finite MPS form.

Global symmetries such as charge and parity conservation manifest as weak symmetries on the Keldysh contour~\cite{buca2012note,albert2014Symmetries}.
For instance, while the system may not have a definite charge due to a thermal initial state, the difference between charges on the forward- and backward branches remains a symmetry of the full dynamics.
We track these weak symmetries in the SGIM by employing symmetric tensor representations with ITensors.jl~\cite{fishman2022ITensor,fishman2022Codebase}.


\paragraph{Automated stitching:} Combining the individually compressed SGIMs and
impurity channels into a single dynamical map requires globally enforcing
fermionic anticommutation. To achieve this, we adopt a Jordan-Wigner (JW)
ordering, where each odd fermionic operator carries a parity string on all
subsequent sites (for alternative approaches see
Refs.~\cite{thoenniss2023Efficient,chen2024Grassmann,sun2025Scalable}). In particular, charge-transporting terms incur a string on all
intermediate sites. Importantly, these parity strings act either on the forward
or the backward Keldysh branch individually. However, for compressed environments,
forward and backward branches are fused, which prevents straightforward
implementation of environment-crossing strings. The latter are often unavoidable, e.g., in propagator calculations where open
strings stretch to the edge of the system as in \cref{fig:unitary_to_im}, or
when the environment and impurity indices of an SGIM are spatially separated as
in \cref{fig:mps_evo}. While it is possible to come up with ad-hoc solutions,
they become rapidly complex and inflexible with more than two environments.

To resolve the generic case, we adopt a superfermion
picture~\cite{schmutz1978Realtime,prosen2008Third,dzhioev2011Superfermion},
where operators acting on the forward and backward Keldysh branches are treated
as two separate fermionic species, with anticommutation enforced between them.
Choosing a JW order that alternates between the two species
(cf.~\cite{brenes2020TensorNetwork}) ensures that parity strings through an
environment act on both branches and, therefore, involve only the weak parity.
After lifting states and operators to the superfermion picture and locally
enforcing the alternating branch ordering, a global JW order is
established via standard manipulations~\cite{corboz2009Fermionic}. To account
for the large SGIM dimension, environment-crossing parity strings are
efficiently implemented with matrix-product operators (MPOs) as illustrated in
\cref{fig:mps_evo}. In App.~\ref{app:autosigns}, we outline an automated scheme
to perform this stitching transparently and efficiently for arbitrary JW
orderings.


\paragraph{Spatial compression:} The dimension of the state space grows
exponentially with the number of environments and fermion species. It is
therefore beneficial to decompose it into a tensor network reflecting the
spatial entanglement structure. Here, we choose an MPS with a separate site for
each environment and each impurity orbital. We take the site ordering to match the JW
order so that all gates act on neighboring sites and evolution proceeds via
TEBD~\cite{vidal2007Classical}.

For the 2IAM, we have an eight-site MPS. Grouping the impurity sites in the
center with two environments on each side as in \cref{fig:mps_evo} allows the
environment and impurity evolution terms to act only on three and four sites, respectively. The remaining choice is which two environments share a
side: those coupled to the same impurity, or those with the same spin.
Empirically, we find that the former, while generating larger total bond
dimensions, decomposes into significantly smaller symmetry sectors and is more
efficient. A truncation threshold $\epsilon=10^{-15}$ yielding moderate bond
dimensions ${\sim} 1000$ is found to be sufficient.


\begin{figure}
    \vspace{-1.5em}
    \phantomsubfloats{
        \phantomsubfloat{\label{fig:spectral_kondo}}
        \phantomsubfloat{\label{fig:spectral_qpt}}
        \phantomsubfloat{\label{fig:tiam_phases}}
        \phantomsubfloat{\label{fig:tstar}}
        \phantomsubfloat{\label{fig:quench}}
        \phantomsubfloat{\label{fig:quench_critical}}
    }
    \includegraphics[width=1\linewidth]{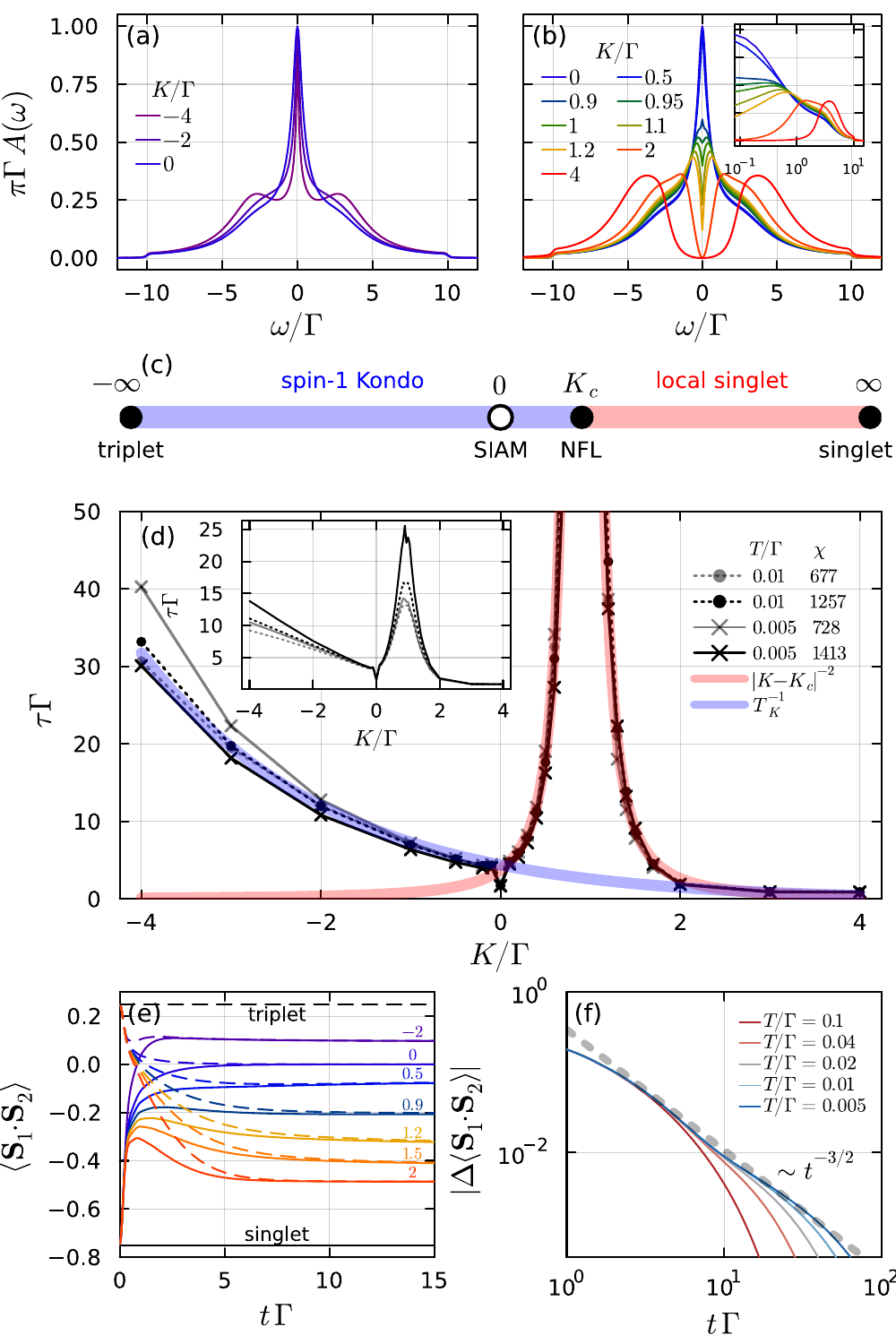}
    \caption{
     Spectral functions for selected $K$ in (a) the spin-1 Kondo regime, and (b) across the quantum phase transition. (c) 2IAM phase diagram with full circles for the fixed points. (d) Finite temperature and SGIM bond dimension collapse to the $T\to 0$ equilibration time after a sudden quench to $K$ with the critical divergence (red) and spin-1 Kondo temperature (blue) superimposed. (e) Quench dynamics from initial impurity singlet (solid) and triplet (dashed) for selected $K$, computed up to $t=100/\Gamma$ but plotted up to $15/\Gamma$ to visibly resolve the decay. (f) Critical quench dynamics at $K=0.9\Gamma\approx K_c$ showing a power-law decay consistent with $t^{-3/2}$ that breaks down to exponential at different temperatures. Unless stated otherwise, $T=0.01\Gamma$, the SGIM bond dimension $\chi=1257$, and the spatial SVD cutoff $\epsilon=10^{-15}$.}
    \label{fig:results}
    \vspace{-1.5em}
\end{figure}

\paragraph{Two-impurity Anderson Model:} As in \cref{fig:model}, 
two spinful impurities $i{=}1,2$ couple to four environments $\alpha{=}i\sigma$
($\sigma{=}\uparrow,\downarrow$), with interimpurity exchange $K$ and intraimpurity Coulomb $U$ interactions at half-filling. The Hamiltonian follows \cref{eq:H,eq:H_Gaussian} with
\begin{equation}\label{eq:TIAM}
    H_\mathrm{imp} = K\mathbf{S}_1{\cdot}\mathbf{S}_2 
    + \sum_{i=1,2} U (n_{i\uparrow}{-}\tfrac{1}{2})(n_{i\downarrow}{-}\tfrac{1}{2}),
\end{equation}
where $n_{i\sigma}=d^\dagger_{i\sigma}d^{\phantom{\dagger}}_{i\sigma}$, $\mathbf{S}_i^{\mu}{=}\frac{1}{2}\!\sum_{\sigma\sigma'}d_{i\sigma}^\dagger\boldsymbol{\sigma}^\mu_{\sigma\sigma'}d_{i\sigma'}^{\phantom{\dagger}}$ and $\boldsymbol{\sigma}^{\mu=x,y,z}$ are the Pauli matrices. Taking a square hybridization\footnote{The band edges are slightly smoothed (cf.~\cite{cohen2015Taming}) to avoid nonuniversal long-time oscillations.} $\Gamma_\alpha(\omega)\approx\Gamma\Theta(D{-}|\omega|)$ in the wide-band regime $D=10\Gamma$, we fix $U=4\Gamma$. 
Figure~\ref{fig:tiam_phases} sketches the phase diagram of the 2IAM, featuring a quantum phase transition from an
interimpurity singlet at $K>K_c>0$ to a conventional spin-1 Kondo phase at $K<K_c$, where the two baths perfectly screen an effective impurity triplet. At the
critical $K_c$, the system forms a non-Fermi-liquid in the same universality
class as the two-channel Kondo effect, characterized by
fractionalization of the impurity degrees of freedom and an effective
Majorana-fermionic description~\cite{zarand2006Quantum,mross2008Twoimpurity,
mitchell2012TwoChannel,sela2011Exact}. Hence, all relevant perturbations, such as deviations of $K$ from $K_c$, have a scaling dimension $\Delta=1/2$. At $K{=}0$, the system decouples into two separate single-impurity
Anderson models, enhancing the Kondo temperature at a single point
(see App.~\ref{app:spin1_Kondo}).

The spectral function $A_\alpha(\omega) =
-\tfrac{1}{\pi}\mathrm{Im}\,G^R_\alpha(\omega)$, here identical for all species $\alpha$, clearly distinguishes the different
 phases. We compute the retarded propagator $G^R_\alpha(\omega) = -i\int_0^\infty\! dt\, e^{i\omega
t}\langle\{d^{\phantom{\dagger}}_\alpha(t),d_\alpha^\dagger(0)\}\rangle$ by first letting the system equilibrate from a quench and then applying the impurity fermionic operators (and JW strings) to the subsequent time evolution.
For $K < 0$ in \cref{fig:spectral_kondo}, we observe the narrowing
of the Kondo resonance at $\omega = 0$ with shrinking $T_K$ as $|K|$ grows. As
$K$ approaches $K_c$ in \cref{fig:spectral_qpt}, the resonance is suppressed
and flattens at low frequencies (inset). Beyond the transition, a gap opens,
signaling the insulating singlet phase.

We now turn to extract the universal exponent of the 2CK point from nonequilibrium protocols. First, the environments are prepared in thermal equilibrium at temperature $T$ and the impurities in a
singlet ($K{\to}\infty$) or triplet ($K{\to}{-}\infty$). Quenching to finite
$K$, we let the system equilibrate and track
$\braket{\mathbf{S}_1{\cdot}\mathbf{S}_2}(t)$ as shown in \cref{fig:quench}. After a nonuniversal transient, it asymptotically relaxes to its thermal value
 as $\braket{\mathbf{S}_1{\cdot}\mathbf{S}_2}(t) {-}
\braket{\mathbf{S}_1{\cdot}\mathbf{S}_2}_\mathrm{eq} {\sim}\ e^{-t/\tau}$.

The relaxation time $\tau$ is cut off by finite temperature $T$ and SGIM bond dimension $\chi$, both of which limit the memory time of each SGIM.
Lower $T$ extends this memory, requiring larger $\chi$. Instead, we make the natural assumption that these cutoffs asymptotically contribute an additive correction to the relaxation rate (cf.~\cite{bayat2016Nonequilibrium}).
Subtracting the (unknown) correction from the numerically computed rate, we recover its genuine zero-temperature value
\begin{equation}\label{eq:tstar_cutoff}
    \tau^{-1}(K) \approx \tau^{-1}(K;T,\chi) - c(T,\chi).
\end{equation}
As $\tau^{-1}(K)$ vanishes at criticality, $c(T,\chi){=}\tau^{-1}(K_c;T,\chi)$. Remarkably, for different but sufficiently low temperatures $T\lesssim 0.01\Gamma$ and moderate $\chi\gtrsim 677$, we find a collapse to $\tau(K)$ over the full $K$ range, as plotted in \cref{fig:tstar}, validating \cref{eq:tstar_cutoff}.
In the Kondo regime, $\tau\sim T_K^{-1}$ tracks the
exponential decrease in the Kondo temperature as $|K|$ grows (see App.~\ref{app:spin1_Kondo}). The suppressed
$\tau$ exactly at $K{=}0$ corresponds to the above-mentioned enhanced Kondo temperature. Approaching $K_c {\approx} 0.91\Gamma$, the system exhibits critical
slowing down, with $\tau$ diverging as $|K{-}K_c|^{-\nu}$ where $\nu =
1/(1{-}\Delta) = 2$.

At the critical 2CK point, the relaxation appears consistent with a power law $\sim t^{-3/2}$, as shown in \cref{fig:quench_critical}. The difference $\Delta\!\braket{\mathbf{S}_1{\cdot}\mathbf{S}_2}\!(t)$ between quenches with an initial impurity singlet and triplet is taken to remove $\braket{\mathbf{S}_1{\cdot}\mathbf{S}_2}_\mathrm{eq}$. To the best of our knowledge,
there is no established theoretical prediction for the scaling of this algebraic relaxation.

\begin{figure}
    \vspace{-1.5em}
    \phantomsubfloats{
        \phantomsubfloat{\label{fig:kz_regimes}}
        \phantomsubfloat{\label{fig:kz_hysteresis}}
        \phantomsubfloat{\label{fig:dissipated_work}}
    }
    \includegraphics[width=1\linewidth]{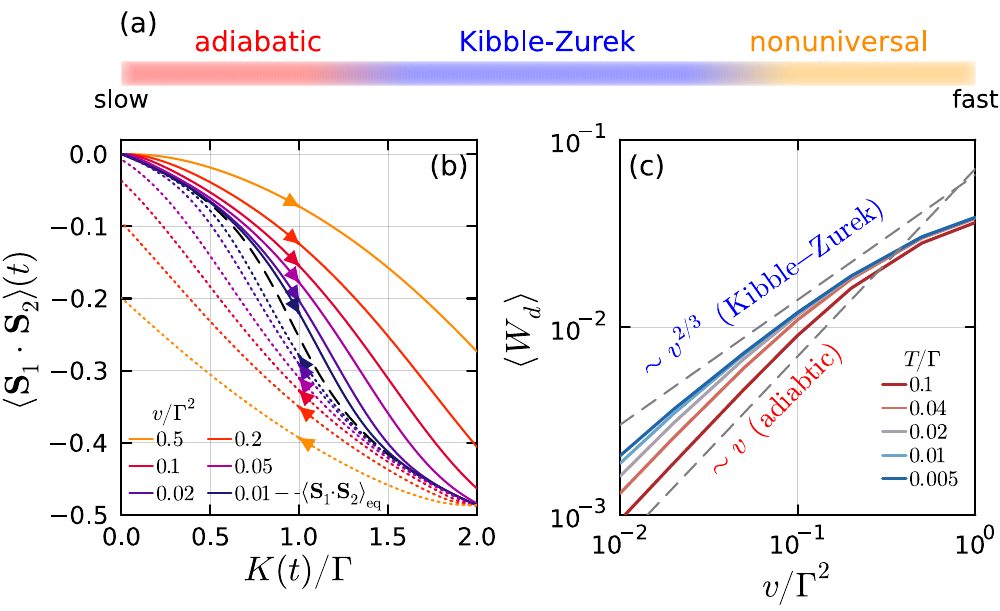}
    \caption{(a) Scaling regimes as a function of the ramp velocity, with the onset of the adiabatic regime dictated by the temperature (or other low-energy cutoff). (b) Interimpurity correlations during ramps of $K(t)=K_0+vt$ across the phase transition, starting from $K_0=0$ (solid) or $K_0=2\Gamma$ (dotted) and varying $v$ at $T=0.01\Gamma$. (c) Dissipated work as a function of $v$ for the forward ramps at different temperatures. Adiabatic and universal scalings are plotted for reference (dashed). Model and simulation parameters are taken as in \cref{fig:results}.}
    \label{fig:kz}
    \vspace{-1.5em}
\end{figure}
Let us now turn to another nonequilibrium protocol to extract the critical exponent.
According to the Kibble-Zurek (KZ)
mechanism~\cite{kibble1976Topology,zurek1985Cosmological},
the divergence of $\tau$ at $K_c$ implies the system must fall out of
equilibrium when a parameter is ramped across the transition at any finite
rate. Here, we ramp $K(t) {=} K_0 {+} vt$ through $K_c$ from equilibrated states on
either side, $K_0 {=} 0$ and $K_0 {=} 2\Gamma$.
While a conventional global-parameter ramp generates topological defects, for an impurity ramp, we quantify the lag behind equilibration by the dissipated work
\begin{equation}\label{eq:Wd}
    \braket{W_d}\equiv W-\Delta F=\!\int_{K_0}^{K_f}\!\!\left[\braket{\mathbf{S}_1{\cdot}\mathbf{S}_2}\!(t)-\braket{\mathbf{S}_1{\cdot}\mathbf{S}_2}_\mathrm{eq}\right]\!dK,\!
\end{equation}
where $t=(K-K_0)/v$ (cf.~\cite{ma2025Quantum,bayat2016Nonequilibrium} for closely
related models in the linear response regime).

Figure~\ref{fig:kz_hysteresis} shows $\braket{\mathbf{S}_1{\cdot}\mathbf{S}_2}$
against the instantaneous $K(t)$ for the forward and backward ramps, with
$\braket{W_d}$ the area between each sweep and the equilibrium curve (dashed).
As $v$ is decreased, the ramp passes through the three regimes sketched in
\cref{fig:kz_regimes}. For fast ramps, the system departs from equilibrium
immediately and $\braket{W_d}$ is nonuniversal. At intermediate velocities, the
system tracks equilibrium far from $K_c$ where $\tau$ is small, but falls out of
equilibrium near the transition as $\tau$ diverges. Since $\tau$ drops on the
other side, the system eventually re-equilibrates, closing a hysteresis loop.
This is the universal KZ regime~\cite{chandran2012KibbleZurek}, where
a scaling ansatz within this out-of-equilibrium window yields $\braket{W_d}\!\sim v^{1/(2-\Delta)}=v^{2/3}$ (see App.~\ref{app:KZ}). For the
slowest ramps, $\tau$ is cut off by finite temperature and bond dimension, and the system stays in the adiabatic regime with $\braket{W_d}\propto v$ by linear
response. \cref{fig:dissipated_work} shows $\braket{W_d}(v)$ for several
temperatures. As expected, at high $T$ the large cutoff drives $\braket{W_d}$
directly from nonuniversal to linear, but for lower $T$ the universal regime
emerges and $\braket{W_d}$ approaches $v^{2/3}$.


\paragraph{Conclusion \& Outlook:}
We introduced an influence-matrix framework for simulating real-time dynamics of multibath impurity models.
It combines accurate per-bath temporal compression into SGIMs and MPS spatial compression while exploiting symmetries, using a flexible stitching procedure to enforce fermionic anticommutation across independently constructed auxiliary baths.
 Applied to the 2IAM, it resolves the spectral function across the
phase diagram as well as the nonuniversal transient quench dynamics. A
single-parameter collapse in bond dimension and temperature recovers the
zero-temperature relaxation rate across the full
parameter range. Near the transition, the relaxation time diverges in agreement with the 2CK non-Fermi-liquid critical exponent. The dissipated work computed across a Kibble-Zurek ramp reveals adiabatic,
2CK-universal, and nonuniversal regimes.

Several directions follow naturally. Multibath SGIM is not tied to any
particular geometry of independent baths. It is equally suited for chains in
which each site is strongly coupled to a
bath~\cite{bruognolo2017Open,fux2023Tensor,bauernfeind2017Fork} and trees for
multi-orbital impurity problems~\cite{cao2021Tree}. Influence matrix approaches
can target dissipative impurity models~\cite{stefanini2025Dissipative,
qu2025Variational, vanhoecke2025Dissipative, sonner2025Semigroup},
nonequilibrium steady states~\cite{chen2024Grassmann,nayak2025Steadystate}, and
driven or interacting
baths~\cite{banuls2009Matrix,lerose2021Influence,giudice2022Temporal}, which now
directly carries over to multibath setups. Moreover, influence matrices have
already proven effective as impurity solvers for
DMFT~\cite{nayak2025Steadystate}, which makes real-time multi-orbital DMFT an
obvious target. The relaxation and ramp protocols studied here are natural
probes in cold-atom realizations of impurity
physics~\cite{bauer2013Realizing,wenz2013Few,riegger2018Localized,hewitt2024Controlling,zhang2020Controlling,amaricci2025Engineering,maki2026Pathway},
while extracting dissipated work and other nonequilibrium thermodynamic
quantities is experimentally within reach in mesoscopic
devices~\cite{saira2012Test,hofmann2016Equilibrium,hofmann2017Heat,barker2022Experimental,han2025Measuring}.
This positions multibath SGIM as a powerful tool for the theoretical support of
experiments realizing strongly correlated nonequilibrium physics.

\begin{acknowledgments}    
\paragraph{Acknowledgments:}
We thank M.~Capone and Y.~Meir for insightful discussions, and particularly
Z.~Ma and E.~Sela for private communication on the KZ
mechanism in impurity models. The work at Princeton is supported by the Brown Investigator Award.
M.~L.~is supported by the Israeli CHE fellowship
for quantum science and technology. M.~S.~is supported by the Swiss National
Science Foundation (SNSF) through a Postdoc.Mobility fellowship (Grant
P500PT\_225372). A.~L.~acknowledges funding through a Leverhulme-Peierls fellowship at the University of Oxford and an Odysseus grant by the Flemish Research Foundation (FWO). This work is partially supported by the European Research Council (ERC) under the European Union's Horizon 2020 research and innovation program (grant agreement No.~864597) (A.~L., D.~A.~A.).
\end{acknowledgments}

\bibliography{misc,main}


\appendix
\setcounter{secnumdepth}{1}

\section{Automated Superfermion Stitching}\label{app:autosigns}
Here we sketch the automatic stitching machinery, using tensor-network diagrammatics.
Following ITensor graphical convention~\cite{fishman2022ITensor}, 
we place arrows on indices to point from kets to bras, for example,
$c\rho = \includegraphics[valign=t]{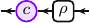}$,
and
$\{c,c^\dagger\} = \includegraphics[valign=t]{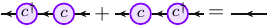} = 1$.
These arrows help to track symmetries and distinguish between the forward and backward Keldysh branches.
For a single site, we define the fermionic operator
$c{=}\sigma^+{=}\left(\begin{smallmatrix}0&1\\0&0\end{smallmatrix}\right)$ and
the associated parity operator
$z{=}1-2c^\dagger c{=}\sigma^z{=}\left(\begin{smallmatrix}1&0\\0&-1\end{smallmatrix}\right)$.
We refer to an operator as odd (even) if it involves an odd (even) number of fermionic
operators.

To enforce anticommutation between fermionic operators on different sites, the Jordan-Wigner transformation attaches semi-infinite parity strings to odd operators. In \cref{eq:jw_superfermions}, these stretch to the right.
We define superfermion operators~\cite{schmutz1978Realtime,prosen2008Third,dzhioev2011Superfermion} by introducing parity strings that act on both branches of the Keldysh contour with alternating order
\begin{subequations}\label{eq:jw_superfermions}
\begin{align}
    c_2^{(\dagger)}\rho&=
    \includegraphics[valign=m]{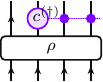}
    \to
    \includegraphics[valign=m]{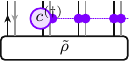}
    =c_{2+}^{(\dagger)}\tilde\rho,\\
    \rho c_2^{(\dagger)}&=
    \includegraphics[valign=m]{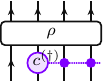}
    \to
    \includegraphics[valign=m]{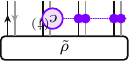}
    =c_{2-}^{(\dagger)}\tilde\rho.
\end{align}
\end{subequations}
Here, purple dots refer to $z$ operators, and the fermionic operator $c_n^{(\dagger)}$ acting on $\rho$
from the left (right) maps to a superfermion operator $c_{n+}^{(\dagger)}$ ($c_{n-}^{(\dagger)}$).
The l.h.s.~guarantees $\{c_n,c_m\}{=}0$ and $\{c_n^{\phantom{\dagger}},c_m^\dagger\}{=}\delta_{nm}$, while the r.h.s.~guarantees the same on each branch,
as well as $\{c_{n-}^{\phantom{\dagger}},c_{m+}^{(\dagger)}\}{=}0$.
We use arrows to distinguish between operators acting from the right and from the left, breaking with the superfermion convention of mapping the bra (on which operators act from the right) to an effective ket with transposed operators.

Operators and states constructed in one JW order can be transformed to a different order by compositions of controlled-Z gates~\cite{corboz2009Fermionic}:
\begin{equation}\label{eq:jwswap}
    \begin{aligned}
        \mathrm{CZ}_{mn} &=1 - \tfrac{1}{2}(1{-}z_m)(1{-}z_n),\\
        \includegraphics[valign=t]{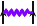} &= \Phi_{\vec{a}\leftrightarrow \vec{a}'} =\prod_{(m,n)} \mathrm{CZ}_{mn},
    \end{aligned}\ \ \ \
     \vcenter{\hbox{\vspace{0.8em}\includegraphics{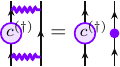}}},
\end{equation}
where $\vec{a}$ and $\vec{a}'$ denote two orderings of the same set of sites and the product defining $\Phi$ runs over every pair $(m,n)$ whose relative order differs between them. As $[\mathrm{CZ}_{mn},\mathrm{CZ}_{kl}]{=}0$ and $\mathrm{CZ}_{mn}^2{=}1$, the global JW-reordering $\Phi$ squares to the identity as well.
Applying $\Phi$ to a wavefunction constructed in one order maps it to a wavefunction in the other order, while operators are conjugated by $\Phi$. This is demonstrated on the r.h.s.~of \cref{eq:jwswap} with $\Phi_{12\leftrightarrow 21}$ mapping a left-stretching to a right-stretching parity string. Note that $\Phi$ can act either in the original or the superfermion picture.
For example, consider an SGIM constructed to have all environment sites on the left. In \cref{fig:mps_evo}, we have such SGIMs, but also SGIMs with the environment to the right. The latter are constructed from a left SGIM by conjugation with $\Phi_{Ei_+i_-\leftrightarrow i_+i_-E}$:
\begin{equation}\label{eq:sgim_flip}
    \vcenter{\hbox{\includegraphics{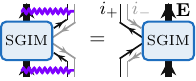}}}\ .
\end{equation}

A key part of the stitching machinery is constructing states and operators in the superfermion picture. Traditionally, superoperators are constructed explicitly by compositions of the r.h.s.~of \cref{eq:jw_superfermions}, while states are constructed by applying operators to the superfermion vacuum. Here, we introduce an alternative approach: states and operators are constructed as tensors in the original picture, i.e., with the l.h.s.~of \cref{eq:jw_superfermions}, and are then automatically lifted to the superfermion picture. 
Consider a typical calculation with arbitrary even operators $L$ and $R$ applied to both sides of an initial density matrix $\rho$, followed by a final trace
\begin{equation}\label{eq:trLrhoR}
    \mathrm{tr}\{\textcolor[RGB]{31,111,191}{L}\rho \textcolor[RGB]{192,57,43}{R}\}\ =\ \  \vcenter{\hbox{\includegraphics{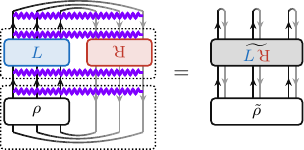}}}\ ,
\end{equation}
where we have geometrically folded the picture to place the backward ($R$, upside down) branch next to the forward ($L$) branch. Inserting identities $\Phi^2$ (in purple) with $\Phi_{1_+\dots N_+N_-\dots 1_-\leftrightarrow 1_+1_-\dots N_+N_-}$ transforming to the alternating JW order yields the lifted state $\tilde{\rho}$ and superoperator $\widetilde{L\rotatebox[origin=c]{180}{$R$}}$.\footnote{Mapping from fermions to superfermions in \cref{eq:jw_superfermions} changes the commutation relations between operators acting on the two branches, and thus cannot be implemented by conjugation. Nevertheless, one can show that composing $\widetilde{L\rotatebox[origin=c]{180}{$R$}}$ (with even $L$ and $R$) and an even number of odd superfermion insertions, applying the result to an initial density matrix, and tracing out (as in spectral-function calculations) is guaranteed to match the original unlifted expression.}
One can show that under this transformation, products $\rho_{\vec{a}}\otimes\rho_{\vec{b}}$ map to lifted products $\tilde\rho_{\vec{a}}\otimes\tilde\rho_{\vec{b}}$, which, in particular, leaves the trace unchanged. This is crucial for factorizing initial states into products of impurity and environment states as done throughout this Letter and is assumed in the definition of the SGIM.

The last piece of machinery consists of combining independently constructed components, each with its own JW order, into a global order. In the global order, the support of each component might no longer be consecutive, as for the SGIM in \cref{fig:mps_evo}. Explicitly, let $O_{\vec{a}\vec{c}}$ be an even operator with regions $\vec{a}$ and $\vec{c}$ forming a consecutive order. We then wish to switch to an order where region $\vec{b}$ comes between them. The resulting operator acts on the joint state space of $\vec{a}$, $\vec{b}$, and $\vec{c}$, and can be represented by three equivalent expressions:
\begin{equation}\label{eq:inject_sites}
    \vcenter{\hbox{\includegraphics{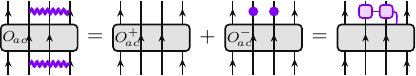}}}\ .
\end{equation}
For concreteness, $\vec{a}{=}(a)$ and $\vec{c}{=}(c)$ are taken as single sites and $\vec{b}{=}(b_1,b_2)$ spans two sites. On the l.h.s.~we insert identities on region $\vec{b}$ and conjugate with $\Phi_{c\, b_1b_2\leftrightarrow b_1b_2 c}$ to directly swap the JW order~\cite{corboz2009Fermionic}.
However, storing and applying the resulting operator is computationally demanding when any of the regions is large, e.g., if $O_{\vec{a}\vec{c}}$ is an SGIM.
Splitting an even $O_{\vec{a}\vec{c}}^{\phantom{\pm}}=O^+_{\vec{a}\vec{c}}+O^-_{\vec{a}\vec{c}}$ 
by the parity of $\vec{c}$ such that $O_{\vec{a}\vec{c}}^{\pm}=\pm z_{\vec{c}}^{\phantom{\pm}} O_{\vec{a}\vec{c}}^{\pm} z_{\vec{c}}^{\phantom{\pm}}$, only $O^-$ carries a parity string across $\vec{b}$. Thus, conditioned on the parity of $\vec{c}$ (or $\vec{a}$, as $O$ is even), on the r.h.s.~of \cref{eq:inject_sites} we attach this string as a bond-dimension-2 parity MPO to $O$, as in \cref{fig:mps_evo}.

Together, the ingredients outlined in this appendix are all that is required to automatically assemble independently constructed SGIMs and impurity channels into a single dynamical map for an arbitrary number of environments and JW orderings.

\section{Kibble-Zurek Scaling}\label{app:KZ}
Here we briefly derive the scaling of the dissipated work~\cite{ma2026private}. The divergence of $\tau\sim|K{-}K_c|^{-\frac{1}{1-\Delta}}$ defines a freeze-out window $|K{-}K_c|< vt_Q$ within which the system is far from equilibrium, with
\begin{equation}\label{eq:tQ}
	t_Q\overset{!}{=}\tau(K=K_c{-}vt_Q)\ \ \Rightarrow\ \  t_Q\sim v^\frac{1}{\Delta-2}.
\end{equation}
The leading contribution to $\braket{W_d}$ comes from this window, and out of it we assume $\braket{\mathbf{S}_1{\cdot}\mathbf{S}_2}(t)$ tracks its equilibrium value. Imposing consistency with the critical scaling out of this window $\braket{\mathbf{S}_1{\cdot}\mathbf{S}_2}_K{-}\braket{\mathbf{S}_1{\cdot}\mathbf{S}_2}_{K_c}\sim\frac{\partial \tau^{-1}}{\partial K}$ \cite{vojta2006Impurity} yields a scaling ansatz~\cite{chandran2012KibbleZurek}
\begin{equation}\label{eq:scaling}
\braket{\mathbf{S}_1{\cdot}\mathbf{S}_2}(t) -\braket{\mathbf{S}_1{\cdot}\mathbf{S}_2}_{K_c}
    =\frac{1}{t_Q^{\Delta}}f\left(\frac{t}{t_Q}\right).
\end{equation}
Restricting \cref{eq:Wd} to the freeze-out window, assuming antisymmetry of $\braket{\mathbf{S}_1{\cdot}\mathbf{S}_2}_\mathrm{eq}$ around $K_c$ to replace it by its critical value (under the integral), and substituting \cref{eq:tQ,eq:scaling}, we arrive at
\begin{equation}
     \braket{W_d}\!\approx\!\!
     \int_{K_c-vt_Q}^{K_c+vt_Q}\!\!\!\underbracket{\braket{\mathbf{S}_1{\cdot}\mathbf{S}_2}\!(t){-}\!\braket{\mathbf{S}_1{\cdot}\mathbf{S}_2}_{\!K_c}}_{\frac{1}{t_Q^\Delta} f\big(\!\frac{K-K_c}{v t_Q}\!\big)}\!\!dK
    \sim v^\frac{1}{2-\Delta}.\!
\end{equation}
A finite temperature (or SGIM dimension) imposes a cutoff on $\tau$ and compromises the restriction to the freeze-out window. The contribution to the dissipated work then comes from the linear-response lag of $\braket{\mathbf{S}_1{\cdot}\mathbf{S}_2}(t)$ behind its equilibrium value, and is linear in $v$~\cite{ma2025Quantum}.

\section{Spin-1 Kondo Temperature}\label{app:spin1_Kondo}
Here we estimate the Kondo temperature in the spin-1 fully screened phase. Let us write an effective spin-1 two-channel Kondo Hamiltonian for \cref{eq:TIAM} in the Schrieffer-Wolff limit~\cite{hewson1997Kondo}, excluding the environment terms
\begin{equation}\label{eq:Heff}
    H \!\overset{K\ll U}{\longrightarrow}\! J\sum_{\!\!i=1,2\!\!}\mathbf{S}_i{\cdot}\mathbf{s}_i + K\,\mathbf{S}_1{\cdot}\mathbf{S}_2
    \!\overset{K\ll 0}{\longrightarrow}\!
    J'\mathbf{S}_\mathrm{tot}{\cdot}(\mathbf{s}_1{+}\mathbf{s}_2).
\end{equation}
In the first step, we assumed the two impurities were decoupled and took the standard single-impurity limit, such that $J=\frac{8D\Gamma}{\pi U/2}$. Then, we projected onto the triplet states, using $\mathbf{S}_i = \tfrac{1}{2}\mathbf{S}_\mathrm{tot}$ under the projection, such that $J'=\frac{J}{2}$. One can also directly take the Schrieffer-Wolff limit with finite $K<0$, arriving at the right-hand side with $J'=\frac{4D\Gamma}{\pi(U/2-K/4)}$. For a $k$-channel model, the Kondo temperature is~\cite{nozieres1980Kondo,gan1993Perturbative}
\begin{equation}\label{eq:TK}
    T_K=\tilde D \, (\rho_0\tilde J)^{k/2}\, e^{-1/\rho_0\tilde J},
\end{equation}
with $\rho_0$ the density of states at the Fermi level, and $\tilde D$ (${\neq}D$) the effective Kondo-model bandwidth. In \cref{fig:tstar} we plot the 2IAM $T_K$ with $\tilde J{=}J'$, $k{=}2$, and $\rho_0{=}1/2D$. Even though we are not deep in the Schrieffer-Wolff regime ($U,|K|\gg \Gamma$), selecting an appropriate $\tilde D$ yields a good match. Equations~\eqref{eq:Heff} and~\eqref{eq:TK} also explain the single point of enhanced $T_K$ at $K=0$, due to the jump in the exponent from $\tilde J=J'\to 2J'$.

\clearpage


\end{document}